\documentclass[11pt,a4paper,oneside,final]{article}
\pdfoutput=1
\usepackage{epsf,amsmath,amssymb,dcolumn}
\usepackage{scalefnt,cite,hyperref}
\usepackage{cleveref,etoolbox,color,soul}
\usepackage{float}
\usepackage{listings}
\usepackage[utf8]{inputenc}
\usepackage{indentfirst}

\usepackage{caption}
\usepackage{subcaption,graphicx}

\Crefname{figure}{Fig.}{Figs.}
\usepackage{placeins}
\usepackage{soul}

\usepackage[normalem]{ulem}

\usepackage[utf8]{inputenc}
\hypersetup{
	unicode,
	pdfproducer={LaTeX},
	pdfcreator={pdflatex}
}

\newcounter{notecount}
\begin{document}
\thispagestyle{empty}

\long\def\symbolfootnote[#1]#2{\begingroup%
\def\thefootnote{\fnsymbol{footnote}}\footnote[#1]{#2}\endgroup}

\vspace{1cm}

\begin{center}
\Large\bf\boldmath
Cosmic muons as PMT and SiPM detector background signals 
\unboldmath
\end{center}
\vspace{0.05cm}

\begin{center}
Gholamreza Fardipour Raki$^{a,b}$\footnote{Corresponding Author, fardipour@ipm.ir},
Mohsen Khakzad$^b$\footnote{mohsen@ipm.ir},
Shehu AbdusSalam$^a$\\[0.4em]
\end{center}

\begin{center}
{\small
{{\sl ${}^a$ Department of Physics, Shahid Beheshti University, Tehran, Iran }\\[0.2em]  
{\sl ${}^b$ School of particles and accelerator, Institute for Research in fundamental sciences (IPM),\\ P.O.Box 19395-5531, Tehran, Iran}}\\[0.4em]
}
\end{center}

\vspace*{1mm}
\begin{abstract}
\noindent
Photomultiplier tube (PMT) and Silicon Photo Multiplier (SiPM) are often used for detecting small number of photons or very weak radiations. A light guide usually connects these photodetectors to the test space. In this article, we investigate the effect of background signals caused by cosmic muons scintillation or interactions with PMT and SiPM, their light guide or input window materials. We study such interactions by making simulations using GATE software package and undertaking experiments using detector circuits developed as in~\cite{Raki:2022lwn}. The background cosmic muons can generate photons which will lead to errors in low radiation and single photon detection experiments especially if standard scintillators are not used. For such experiments, we conclude that the most useful method for cutting down the cosmic muons background should be by conducting the experiments deep underground or inside tunnels with several tens of meters of materials above it.   
\end{abstract}


\setcounter{footnote}{0}

\pagenumbering{arabic}

\section{Introduction}
\label{sec:introduction}

Experiments involving particles, including photon, detections have several sources of background signals. Considering Silicon Photo Multiplier (SiPM) and Photomultiplier tube (PMT) as detectors developed in \cite{Raki:2022lwn}, we address backgrounds signals to such settings due to interactions of the cosmic muon with various parts of the detectors. For instance, undesirable signals can be caused by the scintillation of glass or polymethyl methacrylate (PMMA) in the light guide, epoxy resin, or glass materials in the upper window of the SiPM and PMT input window materials. 

SiPM is a pixelated photodetector of avalanche photodiode (APD), while the single-photon avalanche diode (SPAD) is essentially the single-pixel version. For the SiPM pixels, APDs with integrated resistances in series, are connected in parallel through a metal grid and reverse-biased at a voltage a bit lower than the breakdown voltage. SiPM is made of a two-dimensional array of a large number of sensitive SPADs. The diodes are connected in parallel, and their effect is summed at the output. When particles hit the SiPM, the greater the number of affected diodes at the same time, the higher the peak voltage of the output signal. With forward-biased photodiodes, the greater the number of photons in the diode's frequency sensitivity range, the stronger the signal (due to the reduction in the diode's internal resistance). However, in this way, the effect of a small number of photons is never detectable. For the reversely biased case, the signal strength due to weak radiation depends more on the number of stimulated cells. A cell can be stimulated even by a single photon. Thus by arranging suitable biases, signals due to weak radiations can be amplified to strengths similar to those due to strong radiations~\cite{Raki:2022lwn}. PMT detector's signal depend on the number and energy of incident photons reaching its photocathode. This, in turn, is limited by the input window material transparency to the photon wavelengths. Therefore one should expect a stronger effect of cosmic muon on hitting a SiPM than a PMT system. 

{For low number and single photon detection by SPAD, SiPM, and PMT, the effect of cosmic muons trajectory in the detector is significant. These are used in quantum information-related experiments, dark matter detection, flow cytometry, bio- and chemi-luminescence, astronomical photometry, time-of-flight LiDAR and flash LiDAR~\cite{hadfield2009single, APRILE2012573, ApplicationsSinglePhoton, photonics1998photomultiplier}.} In~\cite{bayat2014scintillation} input window scintillation by X-rays and radioactive sources have been addressed using PMT but not SiPM detector systems. The height of output pulses can be used for removing noise from signals~\cite{photonics1998photomultiplier} collected using PMTs. In this article, our main focus is on the effect of cosmic muons as unwanted signals while using PMT and SiPM. 

The content of this article is organised as follows. In section~\ref{sec:windows} we give a brief overview on the input window and light guide materials used in SiPM and PMT. Section~\ref{sec:Simulations} addresses  the simulations of PMT light guide and input window scintillation due to the passage of energetic muons using GATE software package. This is a simulation platform based on Geant4~\cite{Gate8.0}. In section~\ref{sec:Experiments}, we present experiments concerning the passage of cosmic muons through two PMTs, one equipped without the usual standard scintillator and the other with plastic scintillator BC408. The results are used to compare the effect of the passage of cosmic muons based on the different settings. For these, the electronics developed in~\cite{Raki:2022lwn} were employed. We also performed an experiment to analyse cosmic muon scintillations by SiPM input window with and without a light guide in comparison with the scenarios of having BC408 in addition. The experiments showed that there are stronger background signals due to cosmic muons in SiPM than in PMT because of the nature of the signal production in SiPM which is based on reverse-biased avalanche photodiodes.  

\section{Input windows of SiPM and PMT and light guides}
\label{sec:windows}

High energy photons and charged particles, such as cosmic muons, create photoluminescence (scintillation) photons based on the properties of the PMT or SiPM input window materials. The decay time of the photoluminescence is a crucial factor of the detector systems. In some applications, this glow needs to end more quickly after the primary particle has passed.

The photon absorption characteristic of the window or light guide materials is another important factor for the detector performance. Overlapping the photoluminescence and the absorption spectra, causes a significant part of photoluminescence glows to get absorbed within the same material. In this case, the average path length of the photons before getting absorbed is an important factor. 
In this case the average path length of the photons before getting absorbed in an important factor. Due to the smaller size of SiPM compared to PMT, such photons have more chance of reaching the photodiodes~\cite{ghazy2020preparation}. The emission and absorption spectrum of SiPM input window materials have much more overlap than materials in PMT. 

Fluorescence lifetime or decay time is the duration that the scintillation effect continues producing photons from the time that the energetic particle has gone through the material. The decay time of glass scintillators is around 10 to 15 microseconds\cite{fujimoto2015photoluminescence,kroning2002x}, while the decay time for a plastic scintillator or similar materials is around 2 to 8 nanoseconds\cite{kroning2002x, moser1993principles, chen2022broadband}.

Epoxy resin is a rigid and transparent compound that is used to coat electronic components such as photoreceptors and LEDs, therefore its optical characteristics are crucial. The resin contains three basic compounds: base resin, hardener, and catalyzer. This rigid polymer i.e. the resin which is suitable for the optical application is obtained by combining the three compounds at the proper temperature. Each compound before the combination has special optical characteristics. After the combination process at the proper temperature, polymerisation, and hardening occur, to the effect that the produced polymer will acquire new optical characteristics. For a specific combination, the resin obtains an emission wavelength longer than 300 nm and absorption spectra between 375 to 650 nm. The decay time for this resin is significantly high, around 0.9 seconds\cite{gallot2005identification}.

Polymethyl methacrylate (PMMA) can be precisely cut by infrared (IR) lasers. This type of plastic is also used as a light guide in the detector study which contains SiPM. Although the glass is widely used as a light guide, the decay time of the scintillation is very long compared to the PMMA plastic, around 5 to 10 nanoseconds \cite{ghazy2020preparation}. Comparing the emission and absorption spectra for PMMA, it can be seen that the spectra have a large overlap. So, the produced photons by the passage of an energetic particle, cannot travel more than one centimeter. The peak of the emission spectrum for this kind of PMMA is around 590 nm\cite{ghazy2020preparation}. Undoped PMMA, epoxy resin, and glass luminance efficiencies are very low, and are not transparent to the self-emissions over any significant distance (around 1 cm)\cite{moser1993principles}. However, their emission spectrum is a good match to the common PMTs and SiPMs. Despite the low scintillations, when used within PMT, SiPM, and light guide systems, these materials effectively produce unwanted signals in the weak radiation region due to cosmic muons.

Common scintillators used like BC408 has many characteristics. Those characteristics are optimised to create more photons when an energetic particle goes through the scintillator. The most important one is the wavelength spectrum of emitted photons and the spectrum of photon absorption\cite{Zhang:2020mqz, semiconductorc}. In materials such as simple glass, those two spectra overlap significantly. In the structure of scintillators, the absorption and emission spectra need to be separated to produce a maximum number of scintillation photons. Typical quantum efficiency spectra of common types of PMTs show that the detection efficiency for PMT is between 300 to less than 700 nm\cite{Zhang:2020mqz}. Despite the utterly different structure of the PMT, the SiPM extended to the longer range of absorption. The effective wavelength range for SiPM (MicroFC-30035-SMT) is 300 to 950nm\cite{semiconductorc}.

\subsection{SiPM window}

The window on top of the SiPM sensor is mainly made of epoxy resin instead of glass\cite{MICRORBSERIESD}. For the SiPM MicroFC-60035-SMT, the size of the epoxy resin is about $7\times 7$ mm, and the thickness is 0.21 mm\cite{semiconductorc}. With these dimensions, the emitted photons in the resin have a possibility of reaching the surface of the photodetector cells of the SiPM. It is likely that the mean free path of those photons reaching the photodetector is less than 1 cm, and as a result, they are not absorbed along the path\cite{moser1993principles}. 

\subsection{Input Window and Photocathode of PMT} 
For the case of PMTs, most side-type photomultipliers use a reflection-mode photocathode and a circular structure electron multiplier with high sensitivity and good amplification at a relatively low supply voltage. The head-on type or end-on type has a semi-transparent photocathode, called transmission photocathode, placed on the inner surface of the input window\cite{photonics1998photomultiplier}. Most of the photocathodes used are made of compound semiconductors composed of low-work function alkali metals. The spectral response of the photocathode is expressed according to the type of material.  

The window materials commonly used in photomultiplier tubes are MgF\textsubscript{2} crystal, Sapphire, Synthetic silica and Quartz and Borosilicate glass. The optical properties of these materials can be found at \cite{ghazy2020preparation , nakamura2017scintillation , hamamatsu2007photomultiplier , futami2014optical , croissant2020synthetic , preusser2009quartz , ehrt2009photoluminescence , elkhoshkhany2021investigation}.

\section{Simulations}
\label{sec:Simulations}

\subsection{The simulation of a plastic scintillation}
We use GATE~\cite{Gate8.0} Monte Carlo simulation package based on Geant4\cite{agostinelli2003geant4} for making the simulations related to the studies reported in this article. The implemented geometry is that of a photodetector attached as a PMT, a $40\times 40 \times 10$ mm light guide with different materials -- BC408, glass, epoxy resin, and PMMA -- used over separate simulations. The PMT is attached to the $40\times40$ mm side of the light guide, while the muon source to the opposite with an energy spectrum equivalent to the sea-level cosmic muon radiation~\cite{AUTRAN201877}. By modifying one of the package's cards, "Materials.xml", the characteristics of the scintillation materials were set for the glass, epoxy resin, and PMMA by changing the default values for the BC408 case. The main parameters changed are:
\begin{itemize}
\item{Light Yield (Scintillation Yield)}: Is the number of photons normalised per 1 MeV energy deposited in the scintillator as photons by the trajectory of the energetic particle, in the unit of Photon/MeV. The average light yield has a non-linear dependence on the local energy deposition. Although the light yield of the BC408 is 10000 photon/MeV, in the simulations one can set the value to 1000, because of the following reasons. This is in order to approximate the efficiency of the PMT and significantly increase the speed of the code since there will be relatively fewer optical tracks to compute. In this study, we set the light yield to the least meaningful value\cite{Gate8.0} of 1 Photon/MeV for glass, PMMA, and epoxy.

\item{Fast Time Constant}: Scintillators have a time distribution spectrum with one or more exponential decay time constants, with each decay component having its intrinsic photon emission spectrum. According to the section~\ref{sec:windows}, the decay times for BC408, PMMA and epoxy are set to 2.1 ns. For glass, this is set to 0.9 s.

\item{Fast Component}: This is a histogram that sets the rate of emitting photons at each energy level. This is dependent on the optical properties of the material. In comparison to BC408, the histogram shifts to the higher energies for PMMA, Glass, and epoxy. This means that the emission of PMMA, Glass, and epoxy has higher energy than BC408.

\item{ABSLength}: The photon average path length before getting absorbed in matter. According to the section~\ref{sec:windows}, for BC408 scintillation photons, this parameter is set to 3.8 m. For PMMA and epoxy, it is mostly set to less than 1 cm or to several centimeters for some other wavelengths.
\end{itemize}

\begin{figure}[t!]
  \centering
  \includegraphics[width=0.75\textwidth]{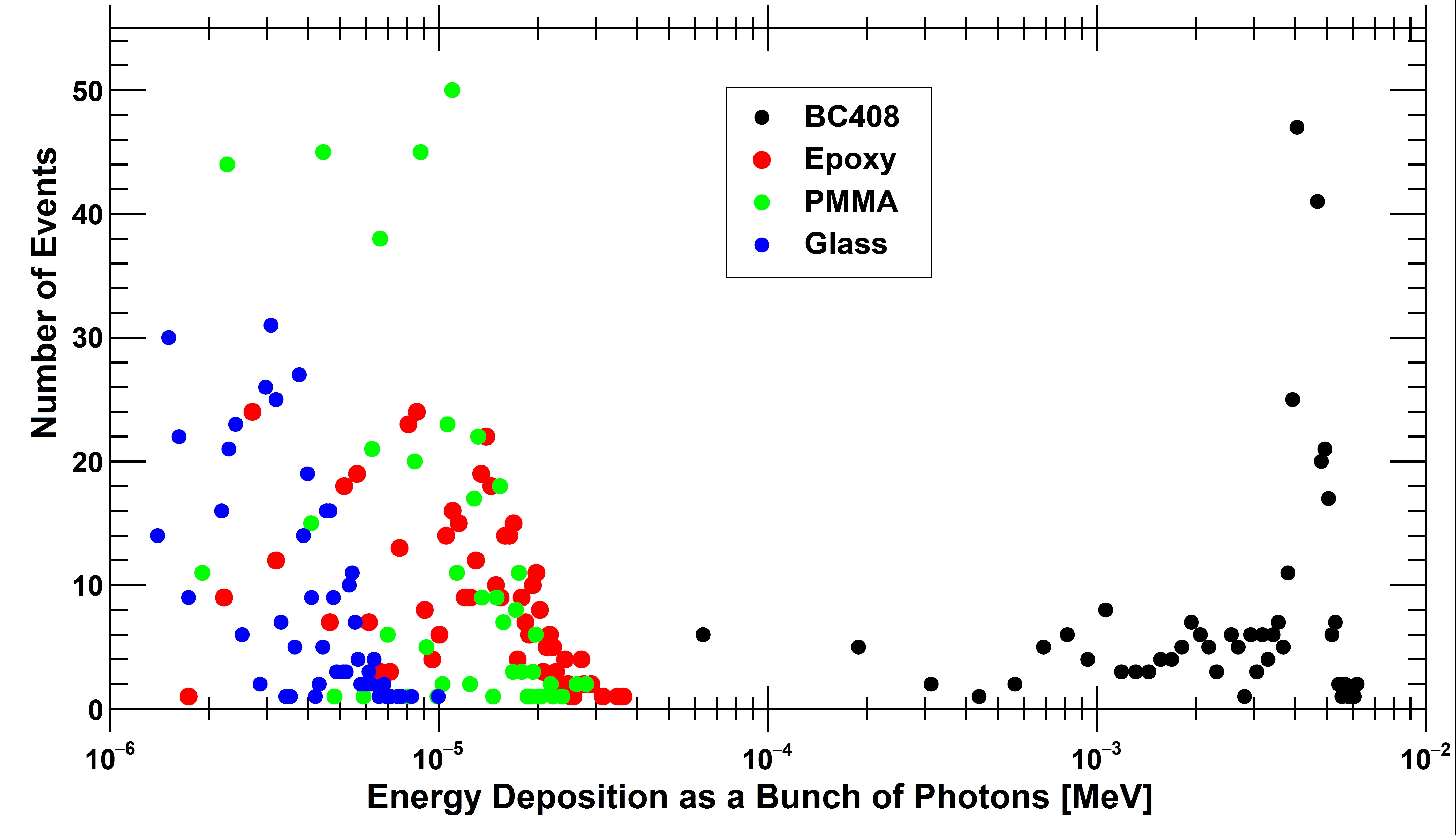}
\caption{The results of the simulation for the number of events done by trajectory of cosmic muons versus scintillation photons bunch energy in the Light guide that is made of PMMA, epoxy resin and glass, detected by PMT.}
\label{fig:simulation}
\end{figure}

Some photodetector surfaces such as PMT are already defined in the GATE, however, for the SiPM depending on the wavelength, it can be adjusted based on the specifications datasheet of the SiPM. According to the datasheet of the PMT, the rate of efficiency for any wavelength can be adjusted as a histogram in the "Surface.xml" file of the GATE package. But Gate8 and Geant4 haven't prepared simulations suitable for the output of SiPM and even PMT, because simulation can only show the energy and number of photos and total energy of photon bunch that arrives at the photodetector surface in each event. Because of the nature of signal production in SiPM by reverse-biased avalanche photodiodes, it can be concluded that the height of the signal caused by weak radiation depends more on the number of stimulated cells and a cell can be stimulated even with one photon. In this way, weak radiations can stimulate a significant number of SiPM cells, and finally, by creating a suitable bias, a signal with a height similar to that of much stronger radiations can be obtained\cite{Raki:2022lwn, AND9770/D}. In PMT the number of photoelectrons, which is equivalent to the height of the signal, is proportional to the number and energy of incident photons\cite{hamamatsu2007photomultiplier}. So the result of the simulation here is more likely to be the PMT output signals than SiPM signals.

Figure~\ref{fig:simulation} shows the number of PMT registered events due to cosmic muons versus scintillation photons generated in the light guide made of PMMA, epoxy resin, and glass. The vertical axis shows the number of muons that collide with the materials. The energy deposited in are shown along the horizontal axis. It shows the magnitude of the energy deposited in the standard scintillation, PMMA, epoxy resin, and glass materials. Given that more deposited energy is equivalent to more number of deposited photons, it can be seen that for every cosmic muon passing through the materials, a number of photons were deposited. Therefore, the simulation shows that the reception of background signals due to the passage of cosmic muons in the detector is certain. The experiments presented in the next section confirm the same result that although the number of signals received for the passage of cosmic muons through the SiPM, PMT detectors are small, but these are significant.

\section{Experiments}
\label{sec:Experiments}

\subsection{Cosmic muons in PMT Hamamatsu R580 input window}
\begin{figure}[t!]
  \centering
  \includegraphics[width=0.4\textwidth]{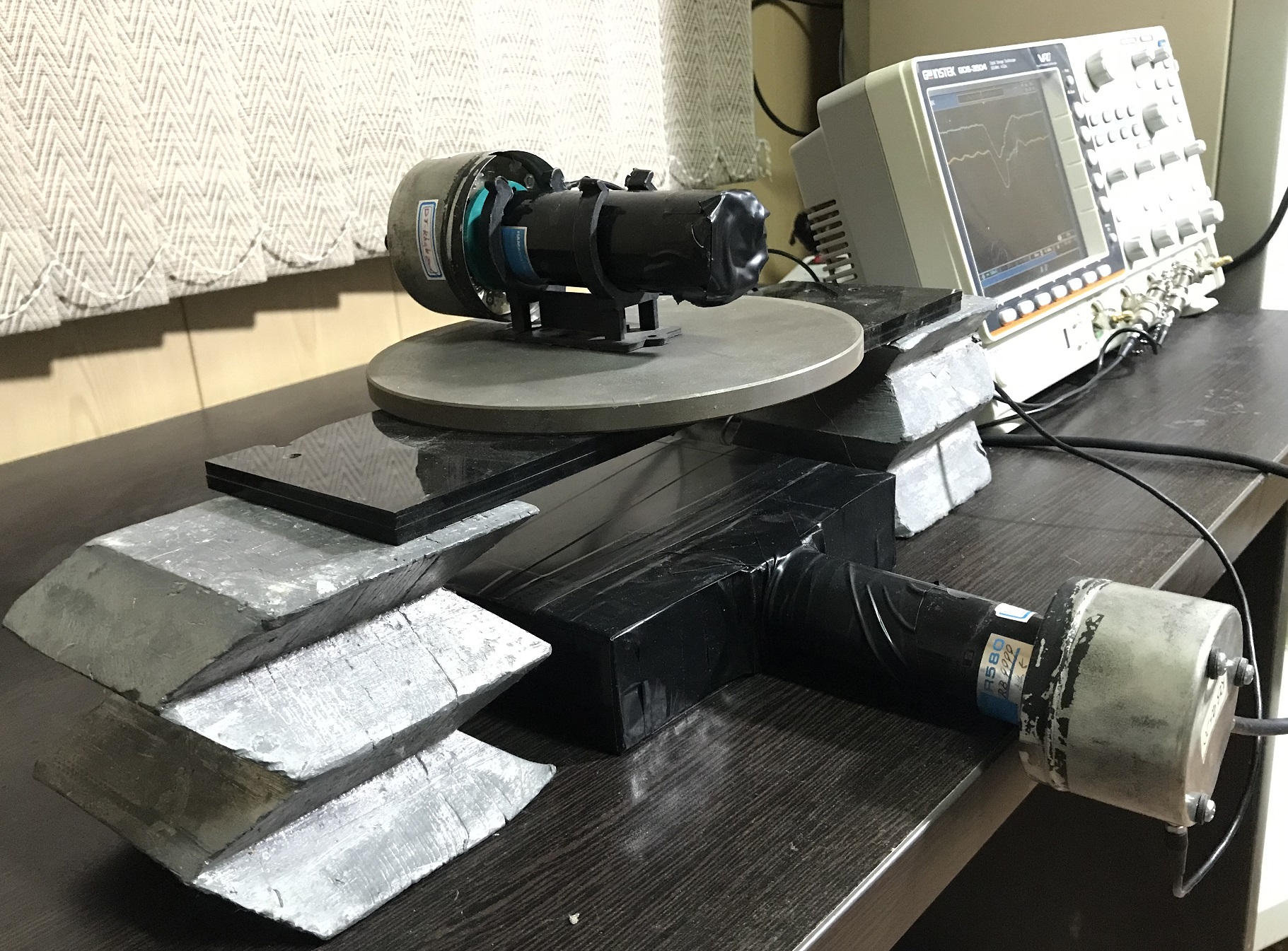}  
  \includegraphics[width=0.5\textwidth]{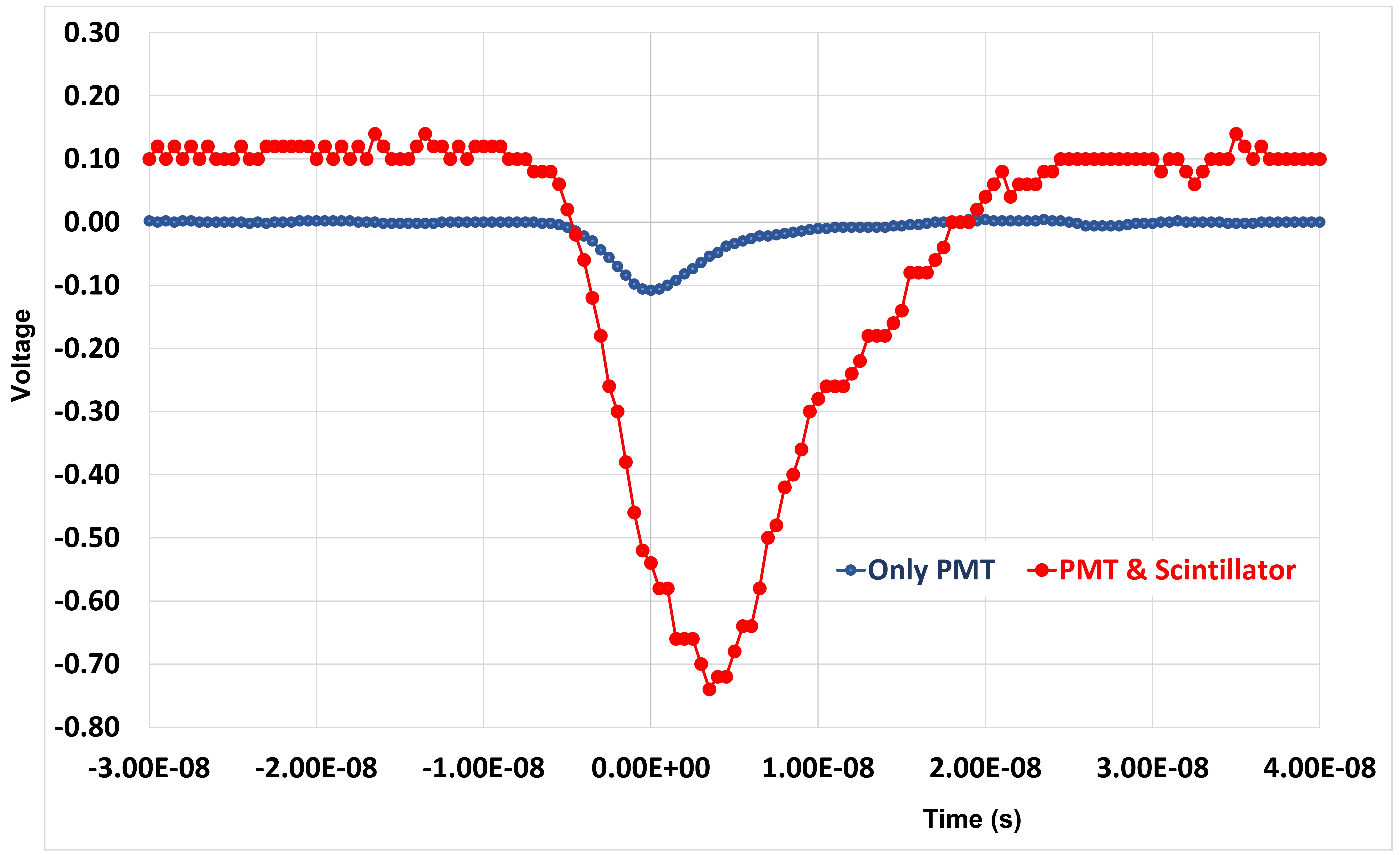}
\caption{(Left) Setup of the experiment of cosmic muon signals in input window of PMT Hamamatsu R580. (Right) A sample set of simultaneous signals made by cosmic muon in the two detectors shown on the left-side of this Figure.}
\label{fig:PMT_ex} 
\end{figure}

R580 is a PMT for scintillation counting and high energy physics, 38 mm (1-1/2 inch) diameter, 10-stage, bialkali photocathode, head-on type. The input window material is Borosilicate glass. In this experiment, we used two similar PMTs, one in the bottom connected to a BC408 plastic scintillator, and the other PMT is not. Between the two detectors there was an iron plate with a thickness of 10 mm to confirm that coincidence signals that may come from two detectors should be caused by a cosmic muon. We used two detectors with completely separate circuits to receive, transmit and amplify the signal. In this way, when signals are received almost simultaneously and with a time difference of a few nanoseconds by both detectors, it can be assured with a high probability that its origin is a cosmic muon passing through both detectors. Being isolated from each other, these two detectors do not induce a signal on each other. Of course, our metal layer between the two detectors reduces the possibility of passing any other type of energetic particle or widespread electromagnetic noise to produce a simultaneous signal in two detectors. Its thickness prevents the passage of particles such as electrons, and its wide metal surface also prevents the passage of electromagnetic noise to some extent. Two detectors were separated by around 10 cm in height. Figure~\ref{fig:PMT_ex}(left) shows the experiment setup.

This test was done on the ground and in a shed. Coincidence signals caused by cosmic muons are two signals that come from two detectors almost at the same time and with a difference of less than 20ns in this setup.  Coincidence signals confirm that a cosmic muon has passed through both detectors. With such a setup, a set of simultaneous signals could be obtained approximately every minute, but by increasing the trigger level, which was set on the signals coming from the above PMT, the number of simultaneous signals was approximately one every five minutes. In this way, it can be seen that on the surface of the earth, a PMT of this type creates a significant background signal caused by cosmic muons on average every minute. A sample set of the coincidence signals from the two detectors are shown in Figure~\ref{fig:PMT_ex}(right).

\subsection{Cosmic muon effect on PMMA light guide and SiPM input window}

\begin{figure}[t!]
\centering
\begin{subfigure}{0.3\textwidth}
\includegraphics[width=1\textwidth]{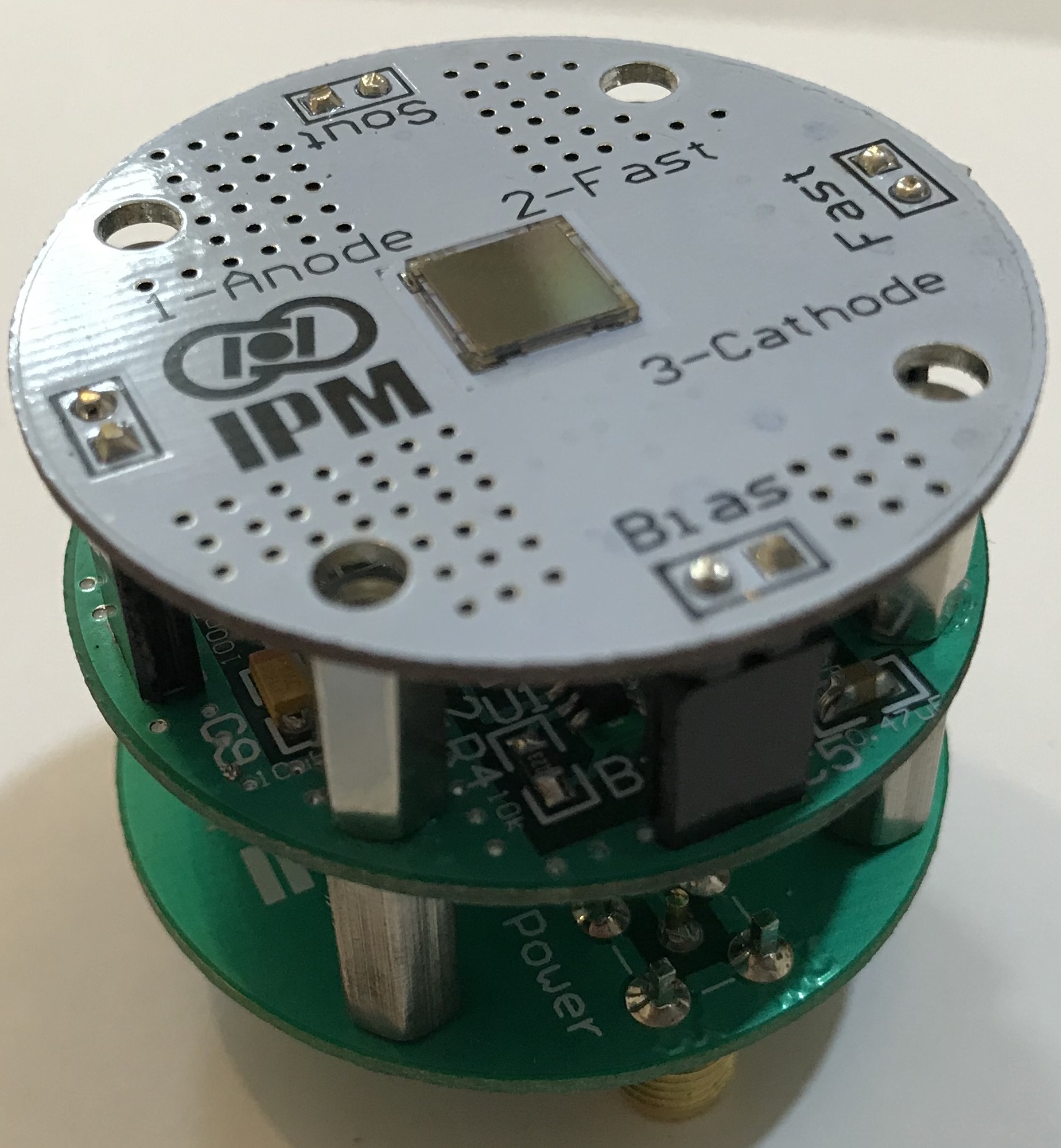} 
\end{subfigure}
\hspace*{2mm}
\begin{subfigure}{0.3\textwidth}
\includegraphics[width=1\textwidth]{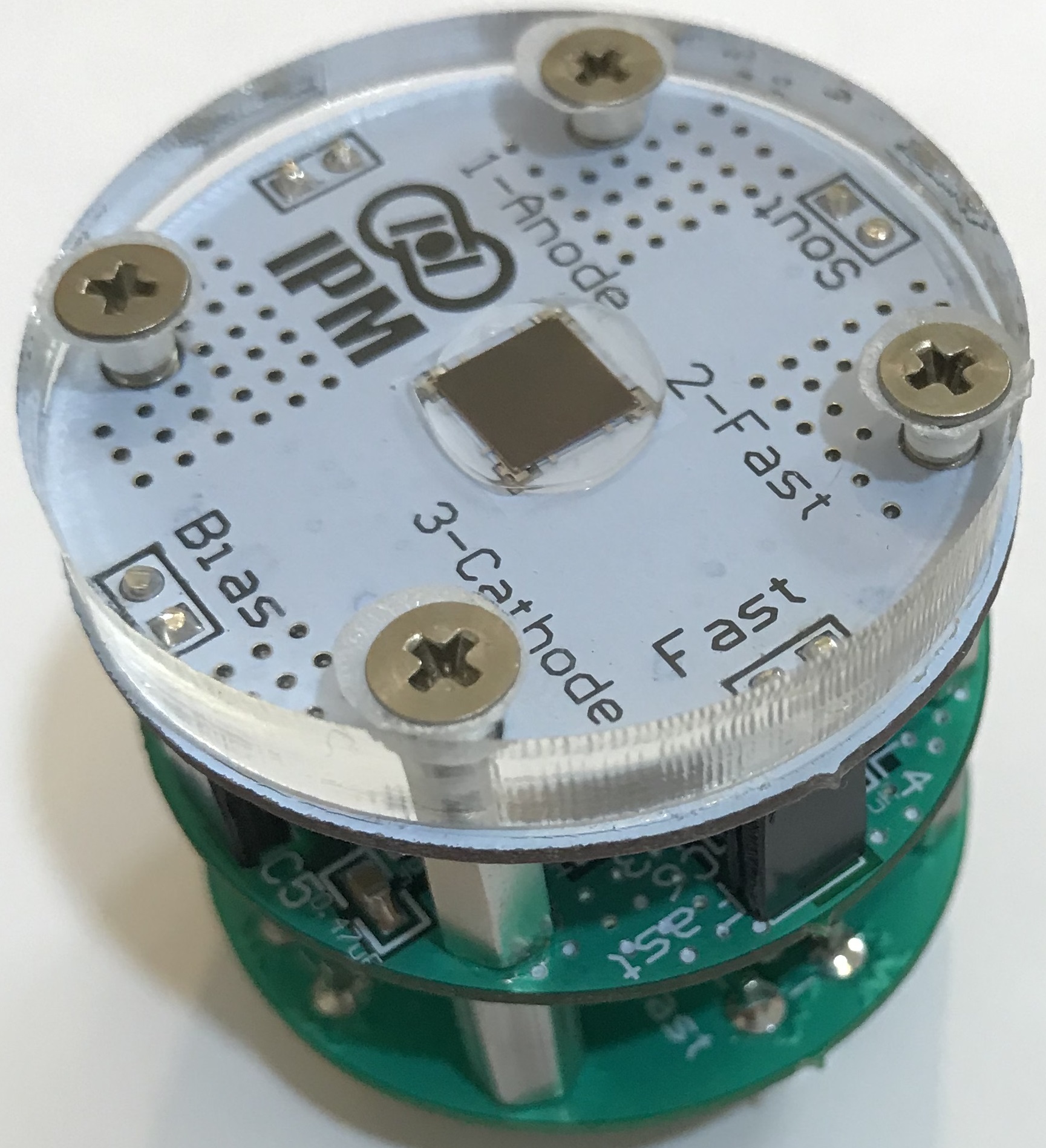} 
\end{subfigure}
\hspace*{2mm}
\begin{subfigure}{0.2\textwidth}
\includegraphics[width=1\textwidth]{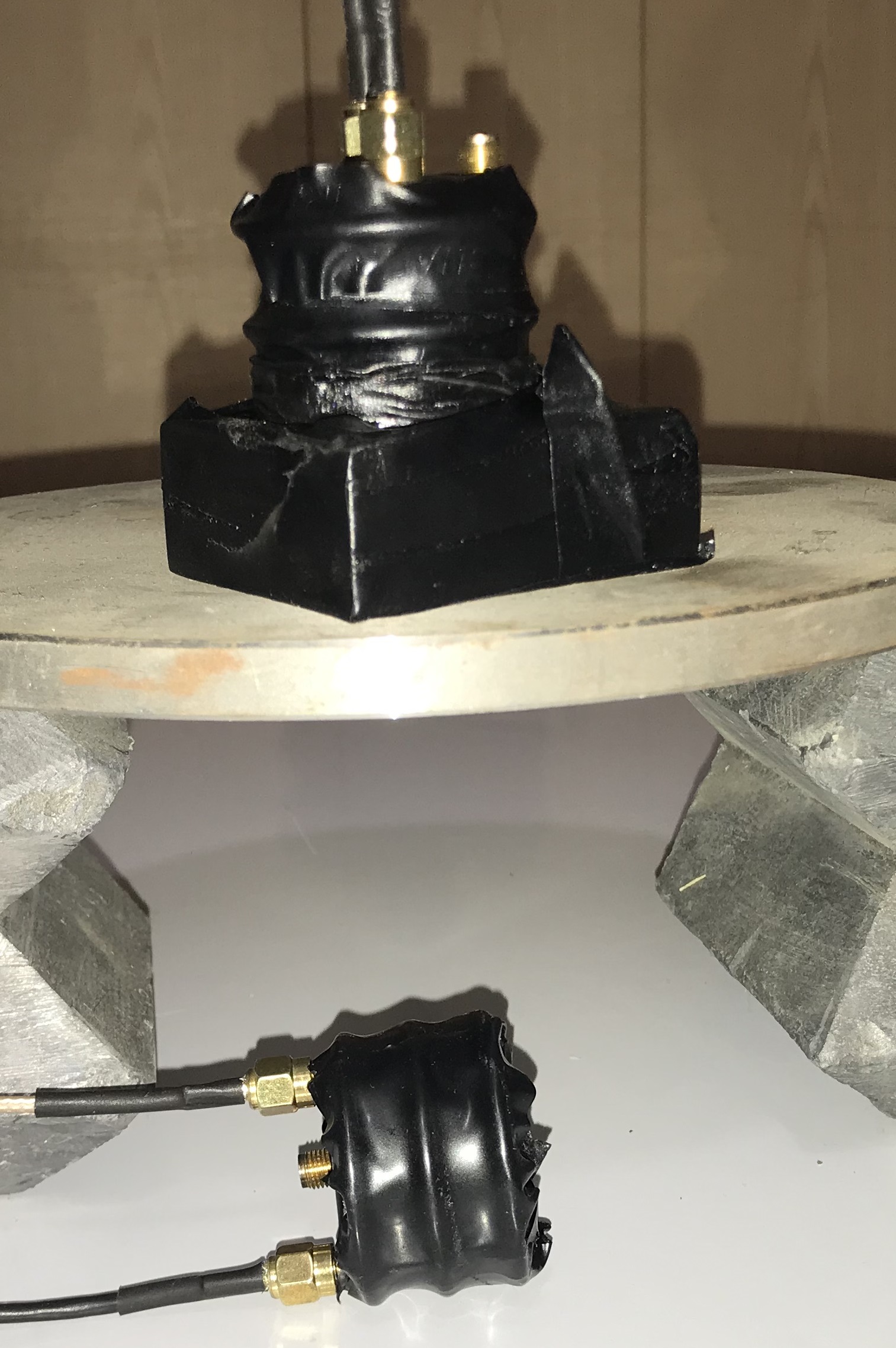}
\end{subfigure}
\caption{(Left) SiPM detector. (Middle) SiPM and one layer of light guide made of PMMA\cite{Raki:2022lwn}. (Right) The structure of the experiment contains one SiPM detector on top that has a light guide and BC408 scintillator and another detector on the bottom, with and without a light guide.}
\label{fig:SiPME}
\end{figure}

This test also was done on the ground and in a shed. We performed a practical scintillation test of the light guide and the SiPM upper window made of epoxy resin using cosmic muons. The flux of cosmic muons on the earth's surface is about one muon per square centimeter per second\cite{Raki:2022lwn}. Each simultaneous signal is obtained for the upper and lower detectors approximately every sixty seconds. We also count the number of simultaneous  signals come from photons due to the effect of muons by the SiPM with no light guide attached to the lower detector. The result was much less than the possibility of receiving every five minutes. Figure~\ref{fig:SiPME} shows the SiPM optical structure of the detector and a light guide made of PMMA as described in~\cite{Raki:2022lwn}. 

\begin{figure}[t!]
  \centering
  \includegraphics[width=0.45\textwidth]{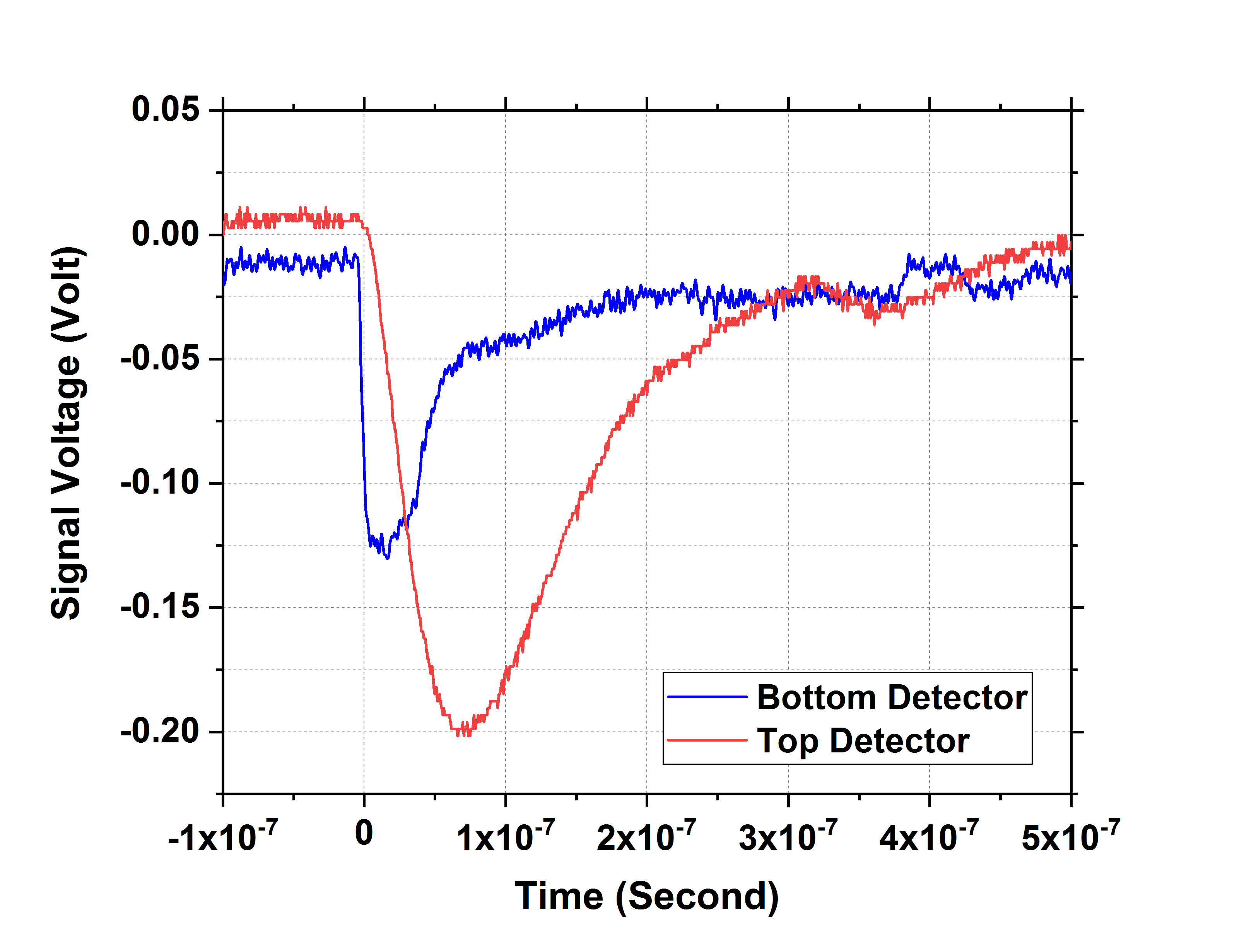}
  \includegraphics[width=0.45\textwidth]{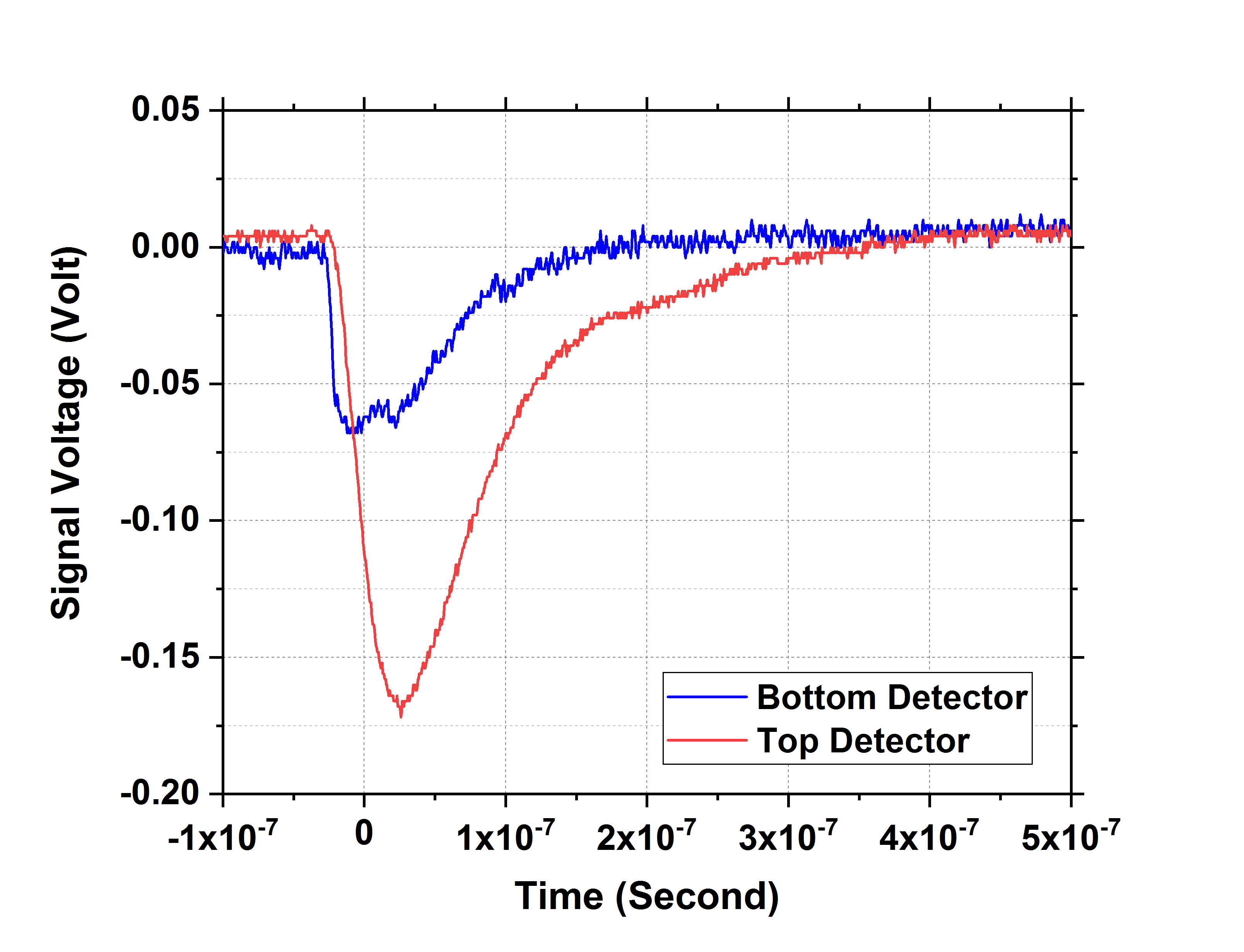}
\caption{(Left) A sample of synchronous signals for the case where the bottom detector (blue line) has a light guide. The top detector (red line) contains light guide and a BC-408 scintillator. (Right) An example of the synchronous signal for the case where the bottom detector (blue line) has no light guide and no scintillation source except SiPM itself. The top detector (red line) contains a light guide and a BC-408 scintillator.}
\label{fig:SiPM_NLG}
\end{figure}

Figure~\ref{fig:SiPME}(right) shows the test setup with a SiPM detector on the top attached to a light guide, BC408, and another detector on the bottom with and without a light guide. When the trigger level that is the voltage level of the acceptance signal in an oscilloscope, moves slowly away from zero value in the oscilloscope, the noise containing the thermal signals in the SiPM is less visible. The simultaneity of the signals obtained by the two detectors confirms a coincidence between them when muons pass through. These same signals confirm that detectors with SiPM, with and without a light guide, generate relatively strong signals compared to thermal noises due to the passage of cosmic muons. A sample of synchronous signals for the case where the lower detector has a light guide is shown in Figure~\ref{fig:SiPM_NLG}(left) and with no light guide in Figure~\ref{fig:SiPM_NLG}(right). The size of the light guide in this detector is equal to the size of the input window of PMT Hamamatsu R580 to make the entrance of these two photodetectors in the same size.

It is expected that PMMA, epoxy resin, or glass scintillations due to cosmic muons be much lower than the case with standard scintillators. The simulations of these materials confirmed the same as shown in Figure~\ref{fig:PMT_ex}(right). For the SiPM, the scintillations due to the input window resin in comparison with cosmic muons are shown in Figure~\ref{fig:SiPM_NLG}. When using SiPM as detector, the number of cosmic muons background hits can be less but with higher effect (signal strength) compared to when PMT is used. This is mostly due to the small dimensions for the former in term of the number of events. All these, together confirm that on the earth's surface in detectors that include PMT or SiPM, the effect of the background signal caused by the passage of cosmic muons is significant, especially for SiPM in terms of height and especially for PMT in terms of number.

\section{Conclusion}
\label{sec:conclusions}
 
We have made simulations and experiments for investigating the effect of cosmic muons on PMT and SiPM detectors. We explicitly show the sensitivities of these to scintillations and interactions with the light guide and input window materials. In comparison, the cosmic muons background signals for SiPM are usually more stronger than for the PMT. However, the smaller size of the SiPM input window will lead to lesser trajectory of cosmic muons over the detector. When the PMT or SiPM is used together with a standard scintillator such as BC408 plastic scintillator, the number of signals caused by cosmic muons in the scintillator is much greater and stronger than the background signal created in the input window and light guides, and therefore these backgrounds are negligible.

\begin{figure}[t!]
\centering
\begin{subfigure}{0.45\textwidth}
\includegraphics[width=1\textwidth]{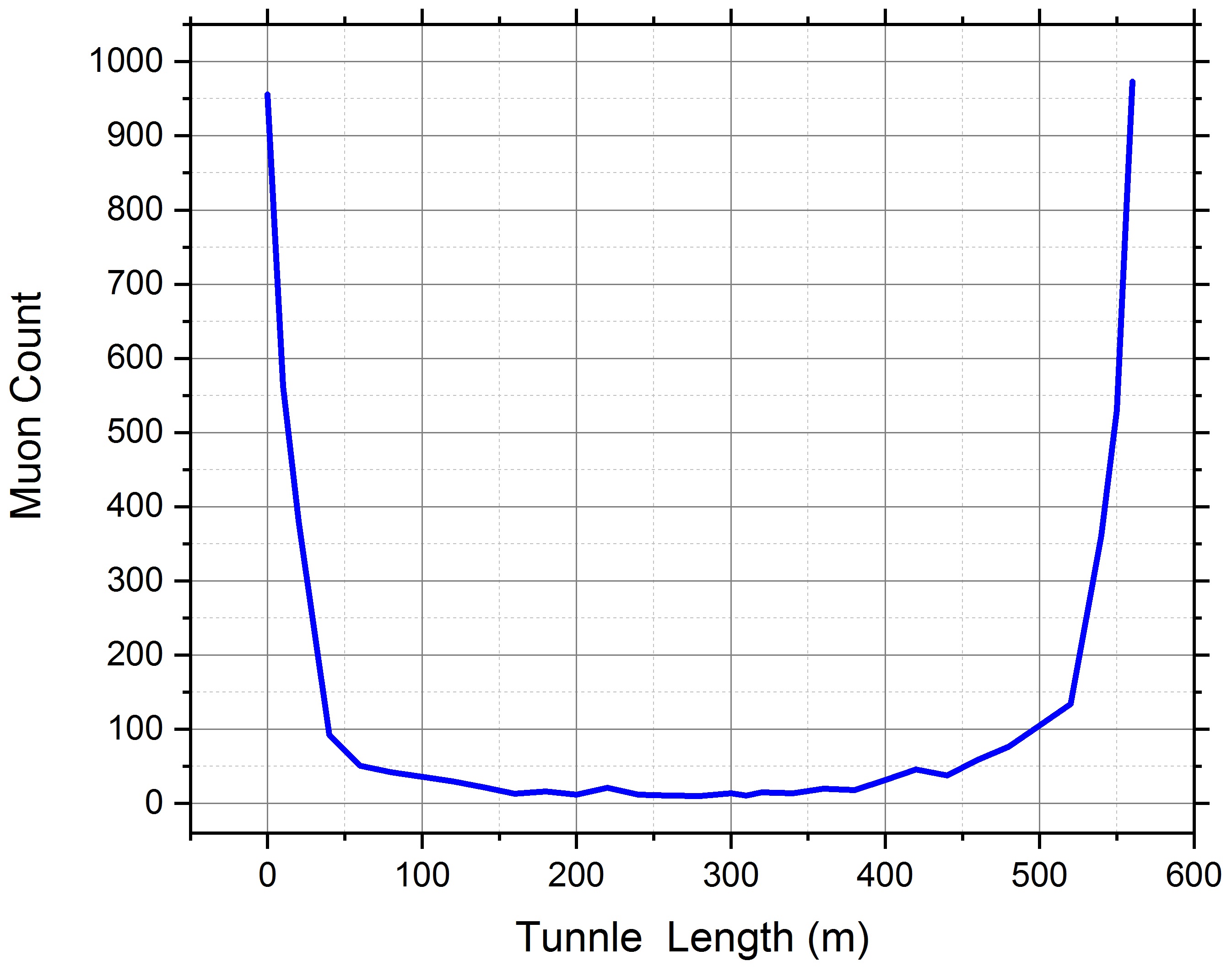} 
\end{subfigure}
\hspace*{2mm}
\begin{subfigure}{0.45\textwidth}
\includegraphics[width=1\textwidth]{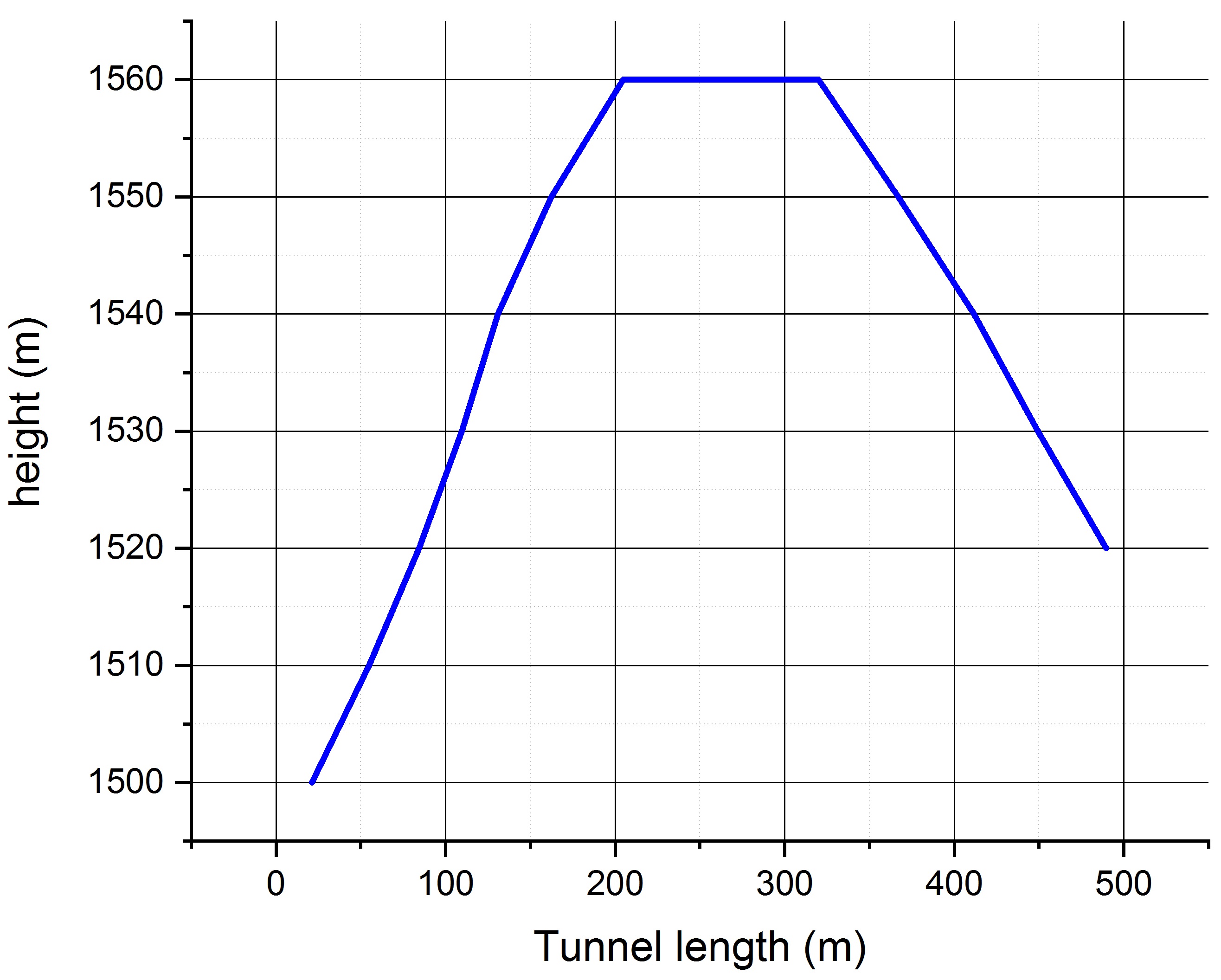}
\end{subfigure}
\vspace*{-3mm}
\caption{The experiment of cosmic muon count in a tunnel. (Left) Muon count in a tunnel with 550m long each count has done in 10 minutes separated by 10 meters. (Right) Curve of the height of matter in the top of the tunnel.}
\label{fig:tunnel}
\end{figure}

The PMT and SiPM are sensitive to cosmic muons. It is therefore important to derive methods for separating such background signals whenever these are used for particles detection. For scenarios where the signals of interest occur at very low rates, such as  single photon and very low radiation detections, the separation of the muon backgrounds is extremely challenging. Based on the studies reported in this article, we conclude that the most useful method for cutting down the cosmic muons background should be by conducting the experiments deep underground or inside tunnels with several tens of meters of materials above. In Figure~\ref{fig:tunnel}, we show the result of cosmic muons counting against distance inside a tunnel. Within 50 meters inside the tunnel, with about 5 meters of rock above, the number of counted muons per 10 minutes decrease from 950 to 50. The number decreases further down to 5 counts per 10 minutes at the centre of the tunnel, with about 60 meters of rock above.

\paragraph{Acknowledgments:} GFR and MK are especially grateful to Dr. Mohammadi Najafabadi from the School of Particles and Accelerators at IPM for the continual interest shown in the project, and the financial support he provided.  

%


{
\bibliographystyle{utphys}
\bibliography{submit.bib}
}

\end{document}